# Abundance of correlated insulating states at fractional fillings of WSe$_2$/WS$_2$ moiré superlattices

Yang Xu[1], Song Liu[2], Daniel A Rhodes[2], Kenji Watanabe[3], Takashi Taniguchi[3], James Hone[2], Veit Elser[4*], Kin Fai Mak[1,4,5*], Jie Shan[1,4,5*]

[1]School of Applied and Engineering Physics, Cornell University, Ithaca, NY, USA
[2]Department of Mechanical Engineering, Columbia University, New York, NY, USA
[3]National Institute for Materials Science, 1-1 Namiki, 305-0044 Tsukuba, Japan
[4]Laboratory of Atomic and Solid State Physics, Cornell University, Ithaca, NY, USA
[5]Kavli Institute at Cornell for Nanoscale Science, Ithaca, NY, USA
Email: ve10@cornell.edu; kinfai.mak@cornell.edu; jie.shan@cornell.edu

**Quantum particles on a lattice with competing long-range interactions are ubiquitous in physics. Transition metal oxides [1,2], layered molecular crystals [3] and trapped ion arrays [4] are a few examples out of many. In the strongly interacting regime, these systems often exhibit a rich variety of quantum many-body ground states that challenge theory [2]. The emergence of transition metal dichalcogenide moiré heterostructures provides a highly controllable platform to study long-range electronic correlations [5–11]. Here we report an observation of nearly two-dozen correlated insulating states at fractional fillings of a WSe$_2$/WS$_2$ moiré heterostructure. The discovery is enabled by a new optical sensing technique that is built on the sensitivity to dielectric environment of the exciton excited states in single-layer semiconductor WSe$_2$. The cascade of insulating states exhibits an energy ordering which is nearly symmetric about filling factor of half electron (or hole) per superlattice site. We propose a series of charge-ordered states at commensurate filling fractions that range from generalized Wigner crystals [7] to charge density waves. Our study lays the groundwork for utilizing moiré superlattices to simulate a wealth of quantum many-body problems that are described by the two-dimensional *t-V* model [3,12,13] or spin models with long-range charge-charge and exchange interactions [14–16].**

Moiré superlattices are formed by stacking two identical lattices with a small twist angle or two lattices with a small period mismatch. The flat electronic minibands afforded by moiré superlattices have led to a plethora of emergent phenomena. Graphene moiré systems exhibit superconductivity, correlated insulating states and topological phases [17–25]. With substantially stronger correlation, transition metal dichalcogenide (TMD) moiré heterostructures manifest a Mott insulator state at one hole per superlattice site and generalized Wigner crystallization of holes at filling factor 1/3 and 2/3 [6,7]. The latter indicates the importance of inter-site (*i.e.* long-range) (*V*) as well as on-site Coulomb repulsion interaction (*U*) when compared to the moiré miniband bandwidth or the kinetic lattice hopping (*t*). As a result of the strong long-range interactions, a rich phase diagram of correlated states is foreseen and calls for experimental and theoretical investigations.

Here we unveil an abundance of correlated insulating states at fractional fillings of a high-quality WSe$_2$/WS$_2$ moiré superlattice. They arise from charge ordering in the underlying superlattice and exhibit a broad range of energy scales ranging from much greater than $t$ to about the same order as $t$. In the limit of $U \gg V$ and $t$ as in WSe$_2$/WS$_2$ moiré superlattices, the system with less than one particle per site is effectively described by the $t$-$V$ model [3,12,13]. Our results show that the excitation energy of the charge-ordered states is widely variable with respect to $t$, and the nature of these states is much richer than the generalized Wigner crystals. This work demonstrates that TMD moiré heterostructures can be used as a quantum simulator of problems described by the $t$-$V$ and related models, potentially including metal-insulator transition, unconventional superconductivity and quantum magnetism [1,2].

The finding is made possible by a new local dielectric sensor in the form of a monolayer TMD semiconductor such as WSe$_2$. The material interacts strongly with light and forms exciton (bound electron-hole pair) states similar to the Rydberg states of a two-dimensional (2D) hydrogen atom [26]. In particular, the exciton excited states (2s, 3s etc.) have Bohr radii many times the monolayer thickness and are sensitive to the dielectric environment [27]. By placing the sensor in close proximity, we probe the insulating states in TMD moiré heterostructures optically from measurements of the resonance energy and oscillator strength of the exciton excited states in the sensor. The technique circumvents the large TMD-metal contact resistance (caused by the large TMD bandgap) that challenges direct electrical transport measurements in TMD moiré heterostructures, particularly, at low carrier densities and low temperatures.

We first demonstrate the technique by probing a WS$_2$ monolayer that can be gate-tuned from an insulating (incompressible) to a metallic (compressible) state. Figure 1a shows the device structure. A WSe$_2$ sensor is placed ~ 1 nm below the sample with a hexagonal boron nitride (hBN) spacer. The spacer thickness is chosen to quench electronic coupling between the sensor and the sample while maintaining their close proximity for high detection sensitivity. The entire structure is encapsulated in ~ 40 nm hBN with a few-layer graphene electrode on both sides. Positive gate voltage $V_g$ is applied between the top graphene electrode and the sample to inject electrons into the sample. The sensor remains charge neutral for the gate range shown.

Figure 1c shows the reflection contrast spectra of the device as a function of $V_g$ at 1.6 K. The prominent features in the three panels from left to right correspond to the 1s, 2s (and 3s) excitons in WSe$_2$ and the 1s exciton in WS$_2$, respectively [26]. For $V_g \gtrsim 1.5$ V (white dashed line), the 1s neutral exciton in WS$_2$ begins to grow faint and the charged excitons emerge at lower energies. This signals electron doping into WS$_2$. The 1s exciton in the sensor, however, shows negligible changes. But the 2s exciton exhibits a stepwise redshift, accompanied by a significant reduction in the reflection contrast. The 3s exciton also becomes much weaker. Two representative spectra corresponding to doped (upper panel) and insulating sample (lower panel) are shown in Fig. 1d. The high-energy half of the spectra is magnified to bring out the 2s and 3s excitons. (See Methods for details on the device fabrication and optical measurements.)

The observed behavior of excitons is similar to earlier studies, in which the medium above the TMD monolayers is changed from vacuum to (conducting) graphene [27]. The phenomenon is explained by renormalization of the exciton binding energies and the quasiparticle bandgap from dielectric screening of the Coulomb interactions by the environment. In the case of the 1s exciton, the two contributions nearly cancel each other and the resonance energy remains unaffected [28]. The 2s exciton emerges as the best probe. When the $WS_2$ monolayer becomes insulating, the 2s exciton binding energy in the sensor increases by about 4 times, and the oscillator strength by 20 times (see Methods). The stepwise gate dependence of the 2s exciton (middle panel in Fig. 1c) is also consistent with the electronic density of states in the 2D sample. The step size is expected to scale with the density of states (see Methods) and to depend on the sample-sensor distance. However, a quantitative description of the data requires comprehensive modeling of nonlocal dynamical screening of the excitonic interactions in 2D TMDs [26,27], which is beyond the scope of this work.

We now apply the exciton probe to examine the insulating states in TMD moiré heterostructures. We replace the $WS_2$ monolayer in the control experiment with an angle-aligned $WSe_2/WS_2$ heterobilayer. Figure 1f is the reflection contrast spectrum of the new device as a function of $V_g$ at 1.6 K. The $WSe_2/WS_2$ heterobilayers have a type-II band alignment: positive $V_g$'s inject electrons into $WS_2$, and negative $V_g$'s inject holes into $WSe_2$ (Fig. 1e). Angle-aligned $WSe_2/WS_2$ heterobilayers form moiré heterostructures with period $a \approx 8$ nm [29] (Fig. 1b). A spectroscopic signature is the formation of moiré excitons (features around 1.68 eV) [6,29–32]. Both the resonance energy and the oscillator strength of the feature exhibit semi-periodic modulations at every half filling of the moiré minibands [6] ($\nu$ = 1, 2 etc. since each miniband has a degeneracy of 2 from the coupled spin-valley degree of freedom [8]). We use positive (negative) filling factor $\nu$ for electron (hole) fillings. The filling factor is initially determined from the ratio of the doping density (evaluated from $V_g$ and the gate capacitance) to the superlattice density (evaluated from the lattice constants and the twist angle between the layers). The spectral range above 1.7 eV is dominated by the sensor response. Similar to the control experiment, the 1s exciton (now centered at 1.725 eV) is doping independent. The 2s exciton (near 1.84 eV) shows very complex doping dependences.

We focus on the 2s exciton in Fig. 2a. The gate dependence exhibits a series of resonance blueshifts, accompanied by an enhancement in reflection contrast. Higher-lying resonances with much smaller oscillator strengths are also observable. They likely arise from the sample-induced moiré bands in the sensor and will not be focused on here. The emergence of the enhanced 2s exciton indicates reduced dielectric screening and opening of a charge gap in the sample. The strongest 2s exciton is observed around 0 V when the sample is charge neutral. The bandgap of the heterostructure is by far the largest energy gap in the system. The next few fillings in descending order of the 2s exciton strength are $|\nu|$ = 1, 2, 1/3 and 2/3. These states have been recently reported in hole-doped $WSe_2/WS_2$ moiré heterostructures with the same energy ordering [6,7]. They correspond to a Mott insulator ($\nu$ = 1), moiré band insulator ($\nu$ = 2), and generalized Wigner crystals ($\nu$ = 1/3, 2/3). Our result shows that these strong insulating states occur on both electron- and hole-filled superlattices, and there are many weaker insulating states, particularly, at fractional fillings.

We refine the gate voltage – filling factor conversion by using the established insulating states as landmarks and assuming a linear dependence for the electron and hole side independently (Extended Data Fig. 1). The two conversion factors are practically identical. The filling factor of the remaining insulating states is determined from the measured $V_g$ as the closest rational number with a small denominator (see Methods). We plot the 2s exciton resonance energy for all observed insulating states with $\nu \neq 0$ in Fig. 2b.

We determine the energy of each insulating state more quantitatively by performing a temperature dependence study (Fig. 3a). Upon heating, the insulating states disappear one by one following a sequence that is largely consistent with the ordering inferred from the 2s resonance energy. We track the monotonic decrease of the 2s spectral weight with temperature (see an example in Extended Data Fig. 4). For each state we estimate the critical temperature $T_C$ from the value at which the spectral weight drops to ~10% of its maximum. Figures 3b-3f illustrate several examples from the electron side with $\nu$ = 1/2, 2/5, 1/3, 1/4, 1/7 and their conjugate states at $1 - \nu$ with the occupied and empty sites switched. The behavior of the $1 - \nu$ state is nearly identical to that of the $\nu$ state. Figure 2c summarizes $T_C$ for all states. The observed 70 K for $\nu = -2$ and 150 K for $\nu = -1$ are in good agreement with earlier electrical transport measurements [6]. Here $T_C$ characterizes the bandgap size for the single-particle moiré band insulator ($|\nu| = 2$) and the thermodynamic phase transition temperature for the other states. Similar analysis can be performed using the 2s resonance energy, but it is less sensitive to temperature than the spectral weight.

The ordering of the $T_C$'s of the insulating states (Fig. 2c) is fully consistent with the ordering of the 2s resonance energies (Fig. 2b). The electron and hole sides behave similarly, suggesting that the same physics is in play. But the interaction effect is stronger on the electron side: the states generally have higher $T_C$'s and a few more states are discernable. In addition, for each side the states at $\nu$ < 1 generally have higher $T_C$'s than the states at $\nu$ > 1. Furthermore, the states at $\nu$ < 1 appear symmetrically about $\nu$ = 1/2 with comparable $T_C$'s for state $\nu$ and $1 - \nu$.

We propose in Fig. 3b-3f the electron configurations on the underlying hexagonal superlattice for states at $\nu$ < 1, including $\nu$ = 1/2, 2/5, 1/3, 1/4, and 1/7. The viability of the configurations were checked, and the charge-ordering temperatures were determined, by performing a classical Monte Carlo simulation with a repulsive Coulomb interaction parameterized by a dielectric constant $\varepsilon$ and gate spacing $d$/2 (see Methods). Using $d$/2 = 40 nm similar to the value in experiment, we find that the model with $\varepsilon$ = 3.9 best describes the highest experimental $T_C$, that of the $\nu$ = 1/3 state. This value is consistent with the dielectric constant of hBN [33] (3, out-of-plane and 6.9, in-plane). That the experimental $T_c$'s of the other states do not all align with the model predictions (Extended data Table 1) we see as evidence that quantum effects cannot be neglected to properly model the system. In the classical model, the conjugate states $\nu$ and $1 - \nu$ share the same $T_C$. Intriguingly, Fig. 3b and 3c show that the $\nu$ = 1/2 and 2/5 (3/5) states have a broken rotational symmetry and are stripe phases. Future experiments are required to verify this.

We briefly discuss the role of quantum fluctuations. In the limit of $t$ = 0, all the insulating states at fractional fillings are generalized Wigner crystals that are described by the configurations shown in Fig. 3b-3f. With increasing $t$, quantum fluctuations can induce charge leakage into the empty sites and turn the Wigner crystals into charge density waves (CDWs) with reduced spatial modulations in charge density. When $t$ becomes comparable to or even larger than the excitation energy of these states, the charge-ordered states melt into metallic states [1]. In WSe$_2$/WS$_2$ moiré superlattices the parameter $t$ is estimated to be ~ 1 meV [8]. If we assume the excitation energy is on the order of $T_C$, we expect $\nu$ = 1/4, 1/3, 1/2, 2/3, 3/4 states on the electron side and $\nu$ = -1/4, -1/3, -2/3 states on the hole side to be close to the generalized Wigner crystal states, and the other states close to the CDWs. In addition, the inclusion of $t$ breaks the symmetry of the energy ordering about $|\nu|$ =1/2 (particularly clear on the hole side). The interaction effect is stronger for $|\nu|$ < 1/2 due to a lower charge density and a larger Coulomb to kinetic energy ratio. The effect of finite $t$ could also explain the observed stronger interaction effects on the electron side. We speculate that the first conduction moiré miniband has a smaller bandwidth than the first valence miniband.

Finally we discuss the insulating states at $\nu$ > 1. The $\nu$ = 4/3 and 5/3 states could be a CDW analogue of the $\nu$ = 1/3 and 2/3 states (the $\nu$ = 1/3 and 2/3 charge configuration plus a completely filled lattice background) or pair density waves if $\nu$ = 1 is a charge transfer insulator [34,35]. The interaction effect is reduced substantially, so is the $T_C$. The $\nu$ = 3/2 state is a peculiar one. Its $T_C$ is significantly higher than that of the $\nu$ = 4/3 and 5/3 states. This is therefore unlikely a CDW analogue of the $\nu$ = 1/2 state. Interestingly, this filling coincides with a van Hove singularity in the density of states of the superlattice [8]. Exotic magnetically ordered states [36,37] and superconductivity [38] have been predicted at van Hove singularities. Whether the $\nu$ = 3/2 state is of exotic origin deserves further investigations.

## Methods
### Device fabrication and electrostatic gating
The van der Waals heterostructures shown in Fig. 1a are prepared by a layer-by-layer dry-transfer method [39] and released on Si substrates (with a 100-nm oxide layer). The raw 2D materials (including the WSe$_2$ and WS$_2$ monolayers and few-layer graphite and hBN) are first mechanically exfoliated from bulk crystals on Si substrates. Flakes of appropriate size and thickness are selected according to their optical contrast. Monolayer WSe$_2$ and WS$_2$ in the moiré superlattice are aligned with a twist angle close to 0 degrees. We employ angle-resolved optical second harmonic generation (details described elsewhere [6,29]) to determine the crystal orientation of the two monolayers and align them accordingly in the stacking process. A thin hBN spacer (typically 1-1.5 nm thick) separates the sample and the WSe$_2$ sensor to quench direct coupling between them. Few-layer graphite is used as contact and gate electrodes. They are connected to the pre-patterned gold electrodes on the substrate. The gate dielectric material hBN on each side has a typical thickness of 35 - 40 nm. To tune the charge density in the sample and keep the sensor layer charge neutral, a voltage $V_g$ is applied to the top-gate electrode through a Keithley 2400 source meter while all other layers are grounded.

### Optical measurements
The reflection contrast spectroscopy measurements are performed with the devices inside a close-cycle optical cryostat (Attocube, attoDRY2100) with variable-temperature capability (1.6 - 300 K). A halogen lamp serves as a white light source, the output of which is first collected by a single-mode fiber and collimated by a 10× objective. The beam is then focused onto the sample by a low-temperature-compatible apochromatic objective with a numerical aperture (N.A.) of 0.8. The beam on the sample has a diameter of about 1 μm and a power below 1 nW. The reflected light from the sample is collected by the same objective and detected by a spectrometer. The reflection contrast ($\Delta R/R_0$) spectrum is obtained by comparing the spectrum of light reflected from the sample ($R$) and from the substrate right next to the sample ($R_0$) as ($R$-$R_0$)/$R_0$. The measurement sensitivity is ~ 0.1%.

### Mechanism of sensitivity of 2D excitons to dielectric environment
Excitons in atomically thin TMDs have been actively studied. Dielectric screening of the Coulomb interactions is known to be important to understanding the exciton binding energy and other key properties [26,27]. The problem of an electron and hole in two dimensions with an attractive $1/r$ Coulomb interaction ($r$ denoting the electron-hole separation) can be solved analytically (the 2D hydrogen model). They form excitons and the binding energy of the $n$-th state is given by $E_b^{(n)} = \frac{m_r e^4}{2\hbar^2 (4\pi\varepsilon\varepsilon_0)^2 (n-\frac{1}{2})^2}$. Here $\varepsilon$ is the dielectric constant of a uniform dielectric medium, in which the exciton is immersed; $\varepsilon_0$ denoting the vacuum permittivity; $m_r$ is the exciton reduced mass; $e$ is the elementary charge; and $\hbar$ is the reduced Planck constant. However, although the bound electron and hole in monolayer TMDs are confined in two dimensions, the electric-field lines between them exist in three dimensions (Fig. 1a). Screening of the Coulomb interaction is thus distance dependent. The screened Coulomb interaction can be well described by the Keldysh potential [40,41]. It scales as ~ $\ln r$ at short-range, and reduces back to ~$1/r$ at long-range. The screening length for monolayer WSe$_2$ in vacuum is ~ 4.5 nm while the Bohr radius of the exciton excited states exceeds 5 - 6 nm [41]. The 2D hydrogen model therefore remains a good approximation for exciton exited states. For instance, the 2s exciton binding energy $E_b^{2s}$ can be extracted from the 2s and 3s energy spacing as $E_b^{2s} \approx \frac{25}{16}(E_b^{2s} - E_b^{3s})$ without knowing the location of the quasiparticle continuum [40,42].

We consider our experimental geometry, where the TMD monolayer (sensor) is in close proximity to another 2D layer with gate-tunable dielectric response. The dielectric response of the medium $\varepsilon(q, E)$ is now a complex function of wavevector ($q$) and frequency or energy ($E$). To correctly account for the dynamic screening effect, for a given $\varepsilon(q, E)$ the exciton binding energy needs to be calculated self-consistently with an energy cutoff determined by the exciton binding energy and a wavevector cutoff by the exciton Bohr radius. Furthermore, the finite spatial separation between the sensor and sample layers needs to be appropriately taken into account. It is challenging to find the exact solution for excitons in this problem. Below we consider the limit of small exciton binding energies to obtain a qualitative picture.

In the limit of small exciton binding energies, we use the dielectric function in the static limit $\varepsilon(q)$. Thomas-Fermi screening yields $\varepsilon(q) = \varepsilon_b + \frac{e^2}{\varepsilon_0 t q^2}\frac{\partial n}{\partial \mu}$ for a uniformly screened Coulomb potential by a free electron gas. Here $\frac{\partial n}{\partial \mu}$ is the electronic density of states or compressibility, $\varepsilon_b$ is the background dielectric response (from bound charges), and $t$ is the thickness of the 2D layer. Combining this result with the 2D hydrogen model ($E_b^{(n)} \propto \frac{1}{\varepsilon^2}$), we expect renormalization in the exciton binding energy in the sensor to be dependent on $\frac{\partial n}{\partial \mu}$ in the sample. Larger density of states would give rise to an enhanced dielectric function and reduced exciton binding energies. This qualitative picture is consistent with our control experiment. The 2s exciton resonance energy shows a stepwise redshift (reduced binding energy) when the sample layer (monolayer $WS_2$) is gate-tuned from an insulating (incompressible) to a metallic (compressible) state. The 2s exciton energy remains constant with further doping since $\frac{\partial n}{\partial \mu}$ is a constant for 2D massive particles.

**Assignment of the filling factor for the insulating states**
The filling factor $\nu$ is defined as the number of electrons or holes per moiré superlattice site. We first estimate $\nu$ from the ratio of the charge density to the superlattice density [6]. The charge density acquired from electrostatic gating is given by the capacitance of the system and applied gate voltage $V_g$. The capacitance is dominated by the geometrical capacitance of the gate and hence is approximately a constant. It is evaluated to be ~ 76 nF/cm$^{-2}$ from the thickness (~35 nm) and out-of-plane dielectric constant (~ 3) of the hBN gate dielectric. The superlattice density is evaluated to be $1.9\times10^{12}$ cm$^{-2}$ from the moiré period for nearly 0-degree aligned $WSe_2/WS_2$ heterobilayer. The uncertainty in $\nu$ is on the order of 10%.

To better determine $\nu$, we use the established insulating state $\nu$ = 1, 2, 1/3 and 2/3 as landmarks and extract the conversion factor from gate voltage to filling factor for electron and hole doping independently. This removes the difficulty with identifying the band edges (i.e. the value of $V_g$ for $v = 0$) and allows a range of $V_g$ for the Fermi level to be inside the heterostructure bandgap. We first determine $V_g$ for all of the insulating states. For states that are clearly identifiable, we obtain their peak position and the full-width-at-half-maximum (FWHM) by performing a Lorentzian fit to the reflection contrast at a fixed photon energy (the 2s resonance energy for the given state) as a function of $V_g$. Extended Data Fig. 4a and 4b show examples of the analysis for state $v = -1/3$ and $-2/3$. For less well-developed states ($1 < v < 2$), we estimate the peak position and peak width from the first derivative of the reflection contrast contour plot (see Extended Data Fig. 2 for an example at 1.6 K).

We perform linear fits using the $V_g$'s for state $v = 1, 2, 1/3$ and $2/3$ on the electron and hole side independently. The slopes are basically identical. They correspond to about 0.25 filling per volt. The value does not depend on temperature strongly up to 50 K. The filling factor of the remaining insulating states is determined from the corresponding gate voltage and the conversion factor. We assign each to the closest rational number with a small denominator since it usually has lower energy. For instance, state $v = 1/4$ is very close to 6/25, 7/30 etc., but is assigned to 1/4 since its denominator is the smallest. Extended Data Fig. 1 summaries the assigned filling factor and the corresponding gate voltage for all observed insulating states. The horizontal bar for each state denotes its FWHM in $V_g$, or equivalently, doping, not the uncertainty in determining the peak position. The uncertainty for the peak position from the Lorentzian fit is substantially smaller.

**Modeling**
In the classical model we keep only the potential energy terms in the Hubbard Hamiltonian:

$$\mathcal{H}(n) = \frac{1}{2}\sum_i \sum_{j \neq i} V(r_{ij}) n_i n_j . \tag{1}$$

Here $n_i \in \{0,1\}$ are the occupations of the sites $i$ of the hexagonal moiré superlattice, $r_{ij}$ is the distance between sites $i$ and $j$, and we use the potential model where the moiré heterobilayer is adjacent to conducting sheets on both sides with spacing $d/2$:

$$V(r) = \frac{e^2}{4\pi\varepsilon\varepsilon_0} \sum_{k=-\infty}^{\infty} \frac{(-1)^k}{\sqrt{r^2 + (k\,d)^2}} \tag{2}$$

We state all lengths in the model in units of the moiré lattice constant $a$ and our unit of energy is $e^2/(4\pi\varepsilon\varepsilon_0 a)$. The potential $V(r)$ decays exponentially with $r$ and the sum $\sum_{j \neq i} V(r_{ij})$ converges to a value independent of the site $i$. This implies, at fixed filling,

$$\mathcal{H}(n) = \mathcal{H}(1-n) + const. \tag{3}$$

Consequently, any thermal equilibrium state at filling $v$ has a counterpart at filling $1 - v$, also in equilibrium.

We located the transition temperatures to the charge-ordered states at fillings $\nu = 1/7, 1/4, 1/3, 2/5, 1/2$ using standard Monte Carlo sampling of the Gibbs distribution for $\mathcal{H}(n)$ using the Metropolis transition rule. The simulations were performed on a periodic $60 \times 60$ hexagonal supercell. To avoid truncation effects, the potential terms $V(r_{ij})$ associated with a site $i$ were replaced by the infinite sum of terms where site $j$ and its occupation is replicated on an infinite hexagonal lattice generated by the supercell. Our transitions maintained $\nu$ and were generated by moving a charge to an unoccupied neighboring site at distance 1 or $\sqrt{3}$. Transition temperatures were determined from the peak in the heat capacity. Because our sampling of the temperature in steps of 0.0005 was too coarse to resolve the first order transitions of the finite (but large) system, the heat capacity peak values in Extended Data Figure 5 should not be given undue significance. However, the precision of the $T_c$'s given in Extended Data Table 1 should be good to within 0.0005.

Our results are for $d/a = 10$ (close to the experimental value). We also simulated $d/a = 5$ and $d/a = 15$, and found that this changed the $T_c$'s by only a few percent. All the heat capacity peaks except for $\nu = 1/3$ are consistent with first-order transitions. The symmetry breaking in the $\nu = 1/3$ state corresponds to the 3-state Potts model, which has a continuous transition. Simulations of the stripe phases were made challenging by the high degree of short-range order in the disordered phase. Very many of our local Monte Carlo transitions were required to mix these non-Wigner-crystal-like systems.

From the fact that even the highest transition temperature ($T_c(1/3) = 0.0695$) is numerically small in our energy unit, we learn that the neglected kinetic energy terms in the Hamiltonian would need to be extremely small for the classical model to be valid. We should therefore not be too concerned that the experimental $T_c$'s of the other fillings, at even lower temperatures, are off the predictions (Extended Data Table 1) based on scaling the model $T_c$ for $\nu = 1/3$.

**References**:
1. Imada, M., Fujimori, A. & Tokura, Y. Metal-insulator transitions. *Rev. Mod. Phys.* **70**, 1039–1263 (1998).
2. Dagotto, E. Complexity in strongly correlated electronic systems. *Science* **309**, 257–262 (2005).
3. Hotta, C. Theories on Frustrated Electrons in Two-Dimensional Organic Solids. *Crystals* **2**, 1155–1200 (2012).
4. Lahaye, T., Menotti, C., Santos, L., Lewenstein, M. & Pfau, T. The physics of dipolar bosonic quantum gases. *Reports Prog. Phys.* **72**, 126401 (2009).
5. Zhu, Q., Tu, M. W. Y., Tong, Q. & Yao, W. Gate tuning from exciton superfluid to quantum anomalous Hall in van der Waals heterobilayer. *Sci. Adv.* **5**, eaau6120 (2019).
6. Tang, Y. *et al.* Simulation of Hubbard model physics in WSe$_2$/WS$_2$ moiré superlattices. *Nature* **579**, 353–358 (2020).
7. Regan, E. C. *et al.* Mott and generalized Wigner crystal states in WSe$_2$/WS$_2$ moiré superlattices. *Nature* **579**, 359–363 (2020).

8. Wu, F., Lovorn, T., Tutuc, E. & Macdonald, A. H. Hubbard Model Physics in Transition Metal Dichalcogenide Moiré Bands. *Phys. Rev. Lett.* **121**, 026402 (2018).
9. Wu, F., Lovorn, T., Tutuc, E., Martin, I. & Macdonald, A. H. Topological Insulators in Twisted Transition Metal Dichalcogenide Homobilayers. *Phys. Rev. Lett.* **122**, 086402 (2019).
10. Wang, L. *et al.* Magic continuum in twisted bilayer $WSe_2$. *arXiv:* **1910.12147** (2019).
11. Shimazaki, Y. *et al.* Strongly correlated electrons and hybrid excitons in a moiré heterostructure. *Nature* **580**, 472–477 (2020).
12. Pietig, R., Bulla, R. & Blawid, S. Reentrant charge order transition in the extended hubbard model. *Phys. Rev. Lett.* **82**, 4046–4049 (1999).
13. Tocchio, L. F., Gros, C., Zhang, X. F. & Eggert, S. Phase diagram of the triangular extended Hubbard model. *Phys. Rev. Lett.* **113**, 246405 (2014).
14. Bak, P. & Bruinsma, R. One-dimensional Ising model and the complete devil's staircase. *Phys. Rev. Lett.* **49**, 249–251 (1982).
15. McKenzie, R. H., Merino, J., Marston, J. B. & Sushkov, O. P. Charge ordering and antiferromagnetic exchange in layered molecular crystals of the θ type. *Phys. Rev. B* **64**, 085109 (2001).
16. Porras, D. & Cirac, J. I. Quantum manipulation of trapped ions in two dimensional coulomb crystals. *Phys. Rev. Lett.* **96**, 250501 (2006).
17. Cao, Y. *et al.* Correlated insulator behaviour at half-filling in magic-angle graphene superlattices. *Nature* **556**, 80–84 (2018).
18. Cao, Y. *et al.* Unconventional superconductivity in magic-angle graphene superlattices. *Nature* **556**, 43–50 (2018).
19. Sharpe, A. L. *et al.* Emergent ferromagnetism near three-quarters filling in twisted bilayer graphene. *Science* **365**, 605–608 (2019).
20. Yankowitz, M. *et al.* Tuning superconductivity in twisted bilayer graphene. *Science* **363**, 1059–1064 (2019).
21. Serlin, M. *et al.* Intrinsic quantized anomalous Hall effect in a moiré heterostructure. *Science* **367**, 900–903 (2020).
22. Lu, X. *et al.* Superconductors, orbital magnets and correlated states in magic-angle bilayer graphene. *Nature* **574**, 653–657 (2019).
23. Chen, G. *et al.* Evidence of a gate-tunable Mott insulator in a trilayer graphene moiré superlattice. *Nat. Phys.* **15**, 237–241 (2019).
24. Chen, G. *et al.* Signatures of tunable superconductivity in a trilayer graphene moiré superlattice. *Nature* **572**, 215–219 (2019).
25. Chen, G. *et al.* Tunable correlated Chern insulator and ferromagnetism in a moiré superlattice. *Nature* **579**, 56–61 (2020).
26. Wang, G. *et al.* Colloquium: Excitons in atomically thin transition metal dichalcogenides. *Rev. Mod. Phys.* **90**, 021001 (2018).
27. Raja, A. *et al.* Coulomb engineering of the bandgap and excitons in two-dimensional materials. *Nat. Commun.* **8**, 15251 (2017).
28. Gao, S., Liang, Y., Spataru, C. D. & Yang, L. Dynamical Excitonic Effects in Doped Two-Dimensional Semiconductors. *Nano Lett.* **16**, 5568–5573 (2016).
29. Jin, C. *et al.* Observation of moiré excitons in $WSe_2$/$WS_2$ heterostructure

superlattices. *Nature* **567**, 76–80 (2019).
30. Tran, K. *et al.* Evidence for moiré excitons in van der Waals heterostructures. *Nature* **567**, 71–75 (2019).
31. Alexeev, E. M. *et al.* Resonantly hybridized excitons in moiré superlattices in van der Waals heterostructures. *Nature* **567**, 81–86 (2019).
32. Seyler, K. L. *et al.* Signatures of moiré-trapped valley excitons in $MoSe_2$/$WSe_2$ heterobilayers. *Nature* **567**, 66–70 (2019).
33. Movva, H. C. P. *et al.* Density-Dependent Quantum Hall States and Zeeman Splitting in Monolayer and Bilayer $WSe_2$. *Phys. Rev. Lett.* **118**, 247701 (2017).
34. Zhang, Y., Yuan, N. F. Q. & Fu, L. Moiré quantum chemistry: charge transfer in transition metal dichalcogenide superlattices. *arXiv:* **1910.14061** (2019).
35. Slagle, K. & Fu, L. Charge Transfer Excitations, Pair Density Waves, and Superconductivity in Moiré Materials. *arXiv:* **2003.13690** (2020).
36. Martin, I. & Batista, C. D. Itinerant electron-driven chiral magnetic ordering and spontaneous quantum hall effect in triangular lattice models. *Phys. Rev. Lett.* **101**, 156402 (2008).
37. Nandkishore, R., Chern, G. W. & Chubukov, A. V. Itinerant half-metal spin-density-wave state on the hexagonal lattice. *Phys. Rev. Lett.* **108**, 227204 (2012).
38. Nandkishore, R., Thomale, R. & Chubukov, A. V. Superconductivity from weak repulsion in hexagonal lattice systems. *Phys. Rev. B* **89**, 144501 (2014).

Extended Data References:

39. Wang, L. *et al.* One-dimensional electrical contact to a two-dimensional material. *Science* **342**, 614–617 (2013).
40. Chernikov, A. *et al.* Exciton binding energy and nonhydrogenic Rydberg series in monolayer $WS_2$. *Phys. Rev. Lett.* **113**, 076802 (2014).
41. Stier, A. V. *et al.* Magnetooptics of Exciton Rydberg States in a Monolayer Semiconductor. *Phys. Rev. Lett.* **120**, 057405 (2018).
42. He, K. *et al.* Tightly bound excitons in monolayer $WSe_2$. *Phys. Rev. Lett.* **113**, 026803 (2014).

**Competing interests:**
The authors declare no competing financial interests.

**Data availability:**
The data that support the plots within this paper, and other findings of this study, are available from the corresponding authors upon reasonable request.

# Figures and figure captions

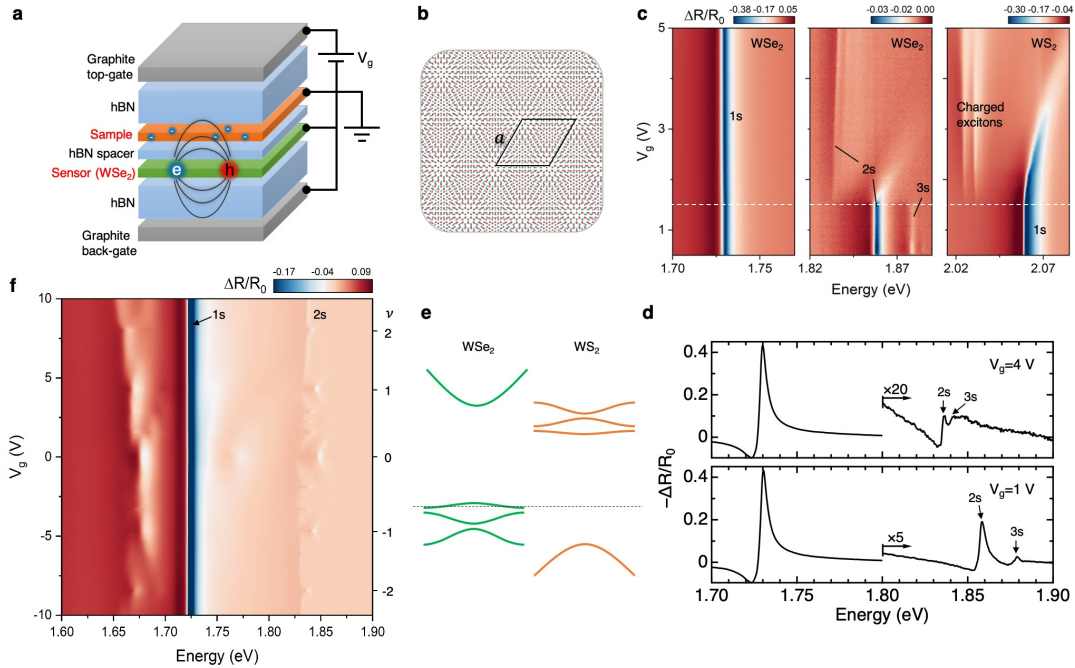

**Figure 1 | Optical sensing of charge gaps using a van der Waals heterostructure platform. a,** Schematic of the device structure and electric circuitry. The bound electron-hole pair (exciton) with its electric-field lines beyond the sensor layer probes the charge gap in the sample controlled by electrostatic gating. **b,** Moiré superlattice of period $a \sim 8$ nm formed by angle-aligned $WSe_2/WS_2$ heterobilayers. Orange and green (blue) circles denote W and Se (S) atoms, respectively. **c,** Gate-dependent reflection contrast ($\Delta R/R_0$) spectra of a control device with a $WS_2$ monolayer as sample at 1.6 K. The three panels (from left to right) show the 1s exciton, 2s (3s) exciton in the sensor, and 1s exciton in the sample, respectively. The sample is electron-doped above $V_g = 1.5$ V (horizontal dashed line). **d,** Two linecuts of **c** at $V_g = 4$ V (top panel) and 1 V (bottom panel). The spectra above 1.8 eV are multiplied by a factor of 20 and 5, respectively. **e,** Schematic band diagram of $WSe_2/WS_2$ moiré heterostructures. The chemical potential (dashed line) can be gate-tuned across the heterostructure bandgap. **f,** Gate-dependent reflection contrast spectrum of a device with a $WSe_2/WS_2$ moiré heterostructure as sample at 1.6 K. The right axis shows the filling factor of the moiré superlattice. The 1s and 2s features are excitons in the sensor.

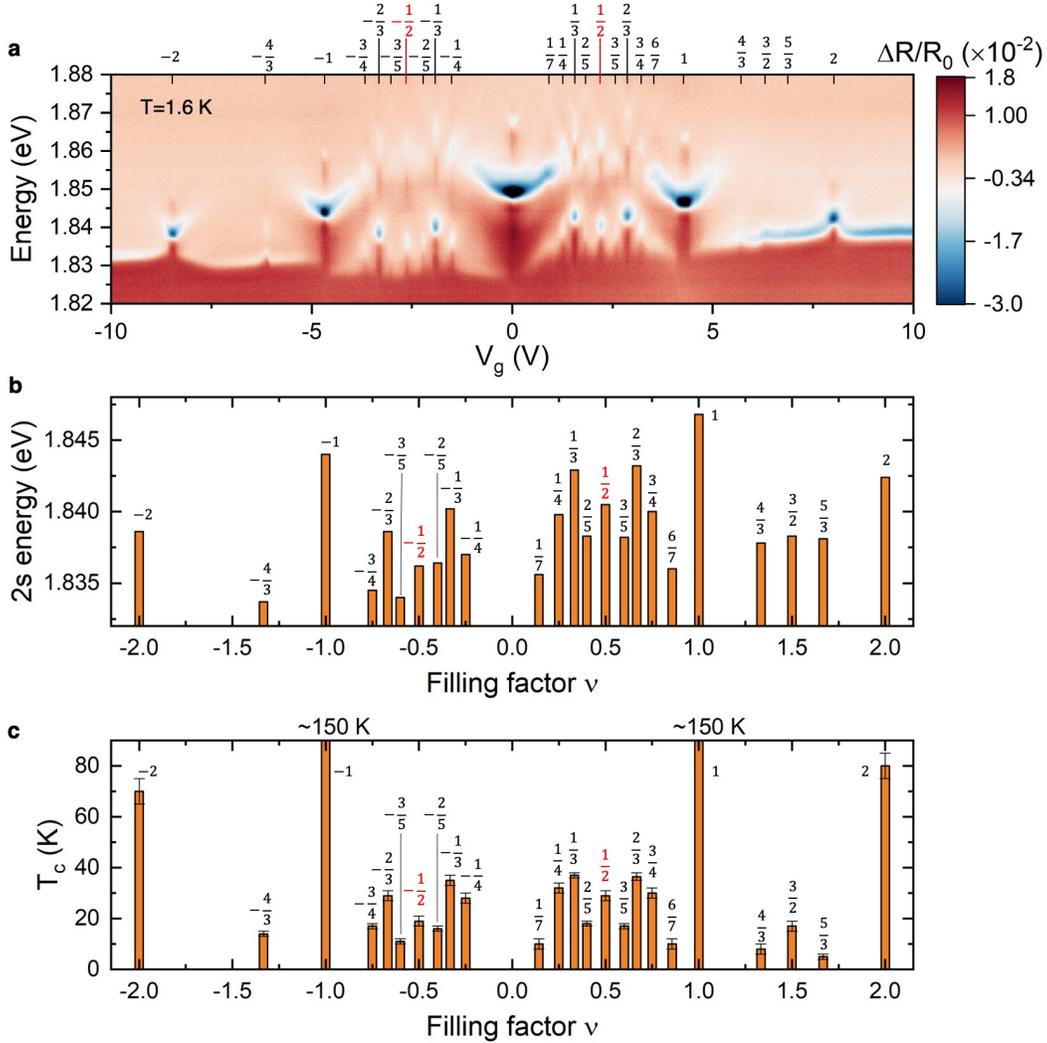

**Figure 2 | An abundance of insulating states and their energy ordering in a WS$_2$/WSe$_2$ moiré heterostructure**. **a,** Detail of Fig. 1f focusing on the 2s exciton in the sensor. An abundance of insulating states is revealed by blueshifts of the 2s exciton resonance, accompanied by an enhancement in the spectral weight. The top axis shows the proposed filling factor for the insulating states. **b, c,** The 2s exciton resonance energy (**b**) and the critical temperature $T_C$ (**c**) for all of the observed insulating states. A uniform width of ~ 0.05 is chosen for all of the states for clarity. It is comparable to the average FWHM of the states. The vertical error bars in **c** are estimated from Fig. 3. For $|\nu| < 1$, the fractional-filling states are symmetric about $|\nu| = \frac{1}{2}$ (marked in red); the energy ($T_C$) of these states is approximately symmetric about $|\nu| = \frac{1}{2}$, particularly, on the electron side.

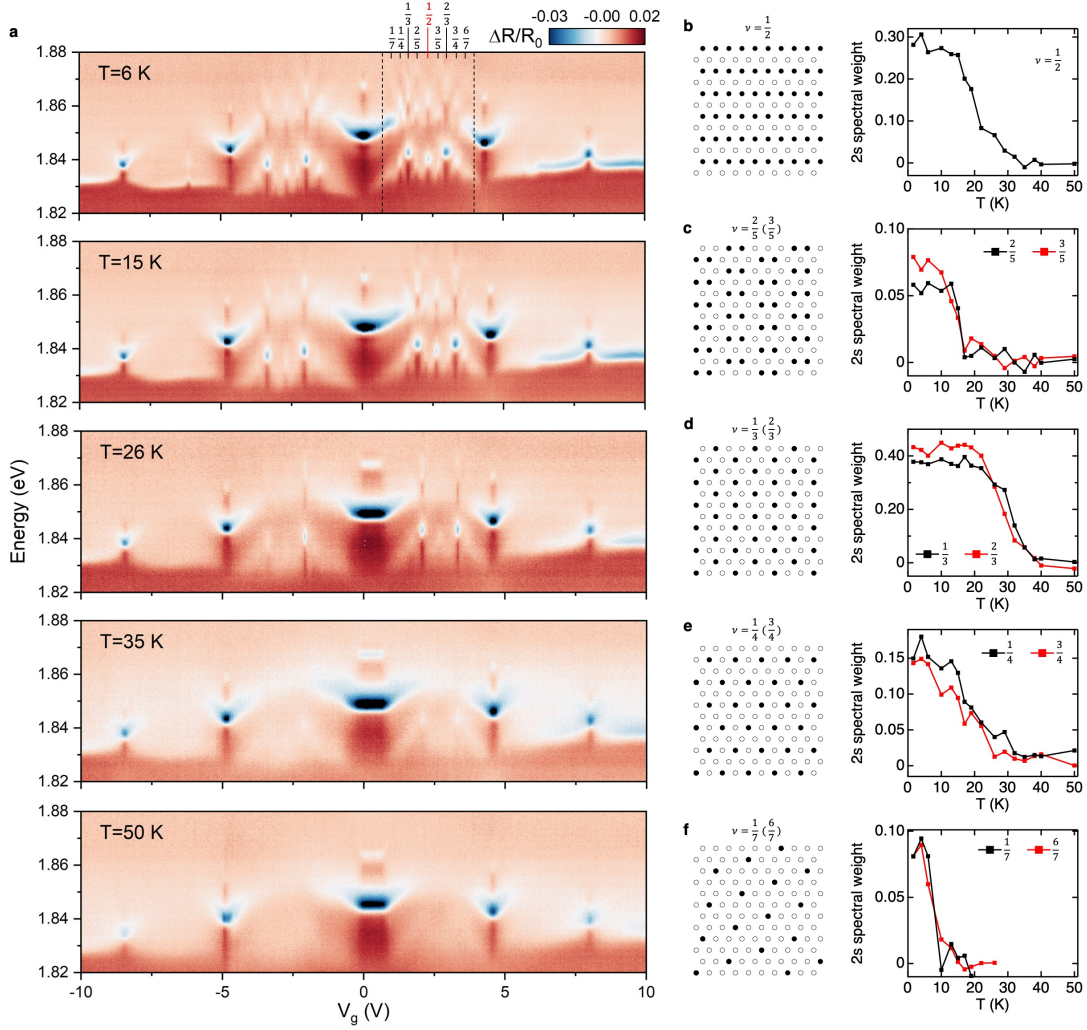

**Figure 3 | Temperature dependence of the correlated insulating states**. **a,** Same as Fig. 2a at elevated temperatures (from top to bottom 6 K, 15 K, 26 K, 35 K, and 50 K). With increasing temperature, the correlated insulating states disappear one by one. **b-f,** Right column: temperature dependence of the 2s exciton spectral weight for state $\nu$ ($=\frac{1}{2}, \frac{2}{5}, \frac{1}{3}, \frac{1}{4}, \frac{1}{7}$) (black symbols) and $1-\nu$ ($=\frac{1}{2}, \frac{3}{5}, \frac{2}{3}, \frac{3}{4}, \frac{6}{7}$) (red symbols). The lines are guides to the eye. Left column: proposed charge-order configuration at zero temperature on the underlying hexagonal moiré superlattice. Filled and unfilled circles denote occupied and empty sites for state $\nu$, respectively. For state $1-\nu$, the notation of the occupied and empty sites is switched.

**Extended data figures and tables**

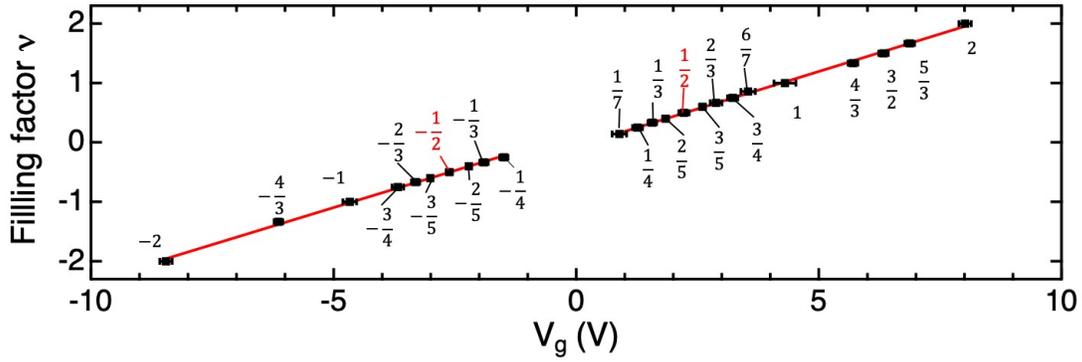

**Extended Data Figure 1 | Assignment of the filling factor of the insulating states.** For each insulating state, the position (filled squares) and the FWHM (horizontal bars) in gate voltage are determined through a Lorentzian fit of the 2s exciton resonance energy as a function of gate voltage. The red lines are linear fits using four established states $\nu = 2, 1, 2/3$, and $1/3$ on the electron and hole sides independently. Both slopes correspond to 0.25 filling per volt. The filling factor of the other states is assigned using the slope as described in Methods.

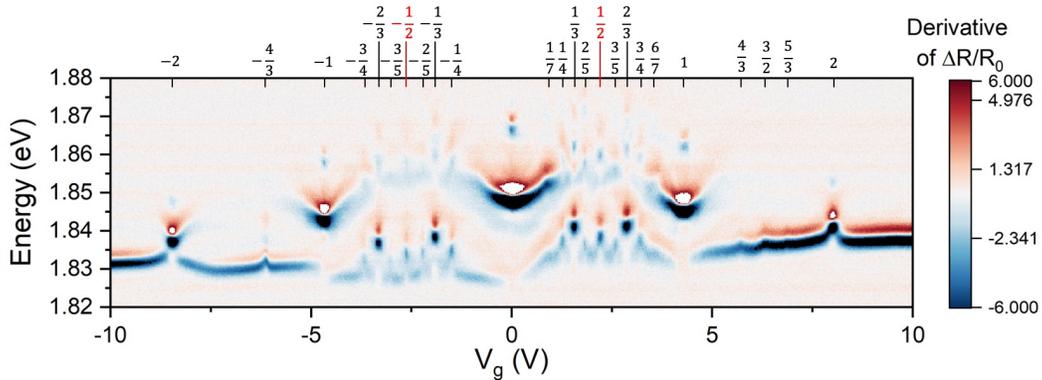

**Extended Data Figure 2 | First derivative of data shown in Fig. 2a with respect to energy.** This is used to evaluate the gate voltage and width of the less well-developed insulating states such as those with $|\nu| > 1$.

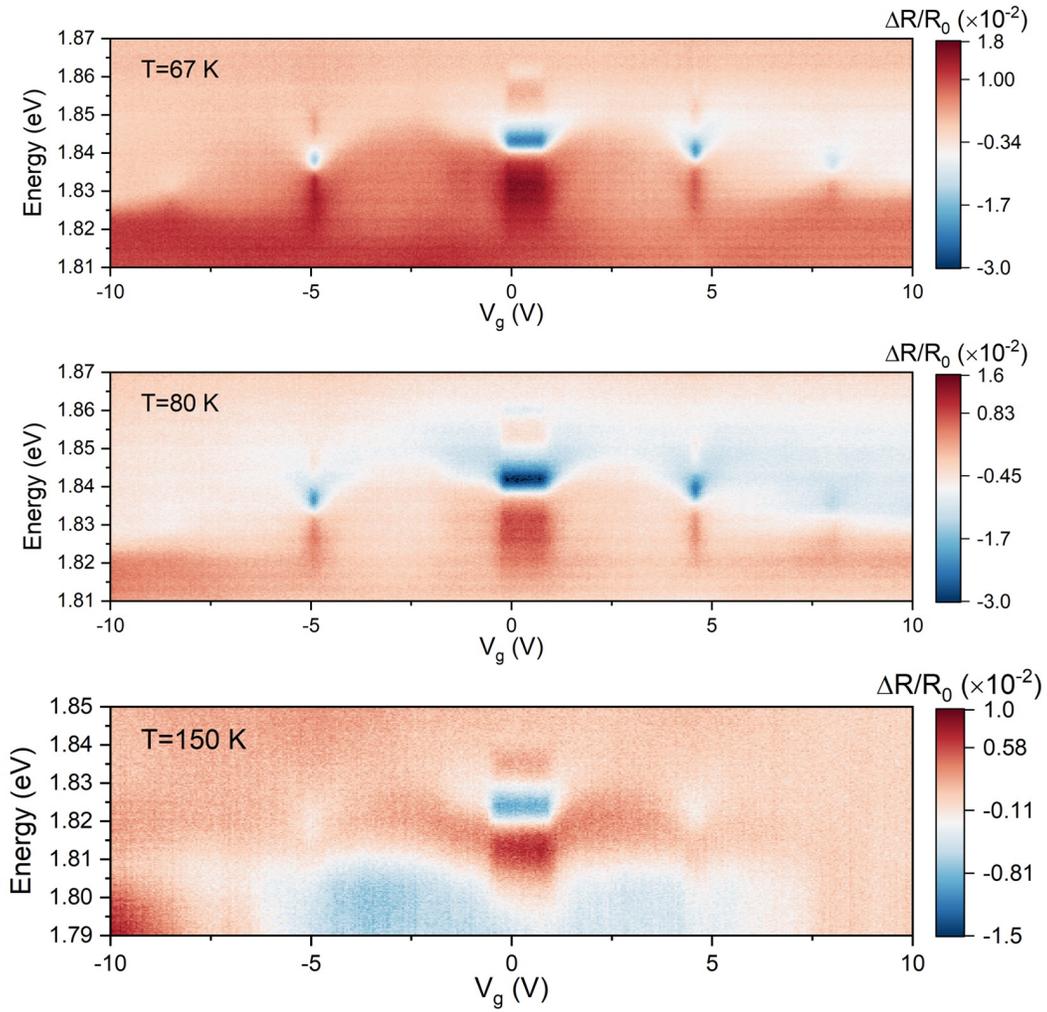

**Extended Data Figure 3 | Contour plots of additional gate-dependent reflection contrast spectrum at higher temperatures (from top to bottom, 67, 80, and 150 K).**

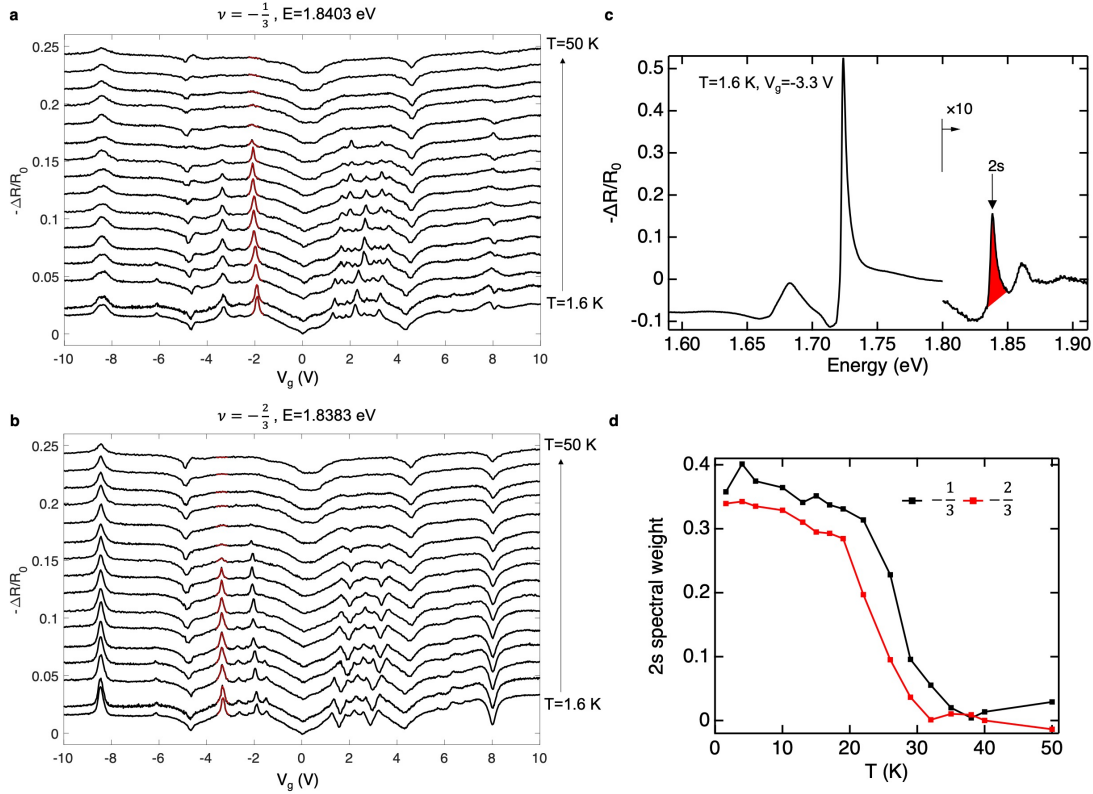

**Extended Data Figure 4 | Analysis of the -1/3 and -2/3 states. a, b,** Two horizontal line cuts of Fig. 2a at 1.8403 eV (**a**) and 1.8383 eV (**b**). These energies correspond to the 2s exciton peak energy for the $\nu = -\frac{1}{3}$ and $\nu = -\frac{2}{3}$ states, respectively. They appear as peaks in $(-\Delta R/R_0)$. Results at different temperatures (in ascending order from bottom to top, 1.6 K, 4 K, 6 K, 10 K, 13 K, 15 K, 17 K, 19 K, 22 K, 26 K, 29 K, 32 K, 35 K, 38 K, 40 K, and 50 K) are vertically displaced for clarity. The red curves are Lorentzian fits to the peaks for the corresponding states. **c,** Reflection contrast spectrum $(-\Delta R/R_0)$ for $V_g = -3.3$ V ($\nu = -\frac{2}{3}$) at 1.6 K. The red area underneath the 2s peak is integrated to obtain the 2s spectral weight. **d,** The 2s spectral weight as a function of temperature for state $\nu = -\frac{1}{3}$ (black symbols) and $\nu = -\frac{2}{3}$ (red symbols). The $\nu = -\frac{1}{3}$ state has a slightly higher $T_C$ than the $\nu = -\frac{2}{3}$ state. The lines are guides to the eye.

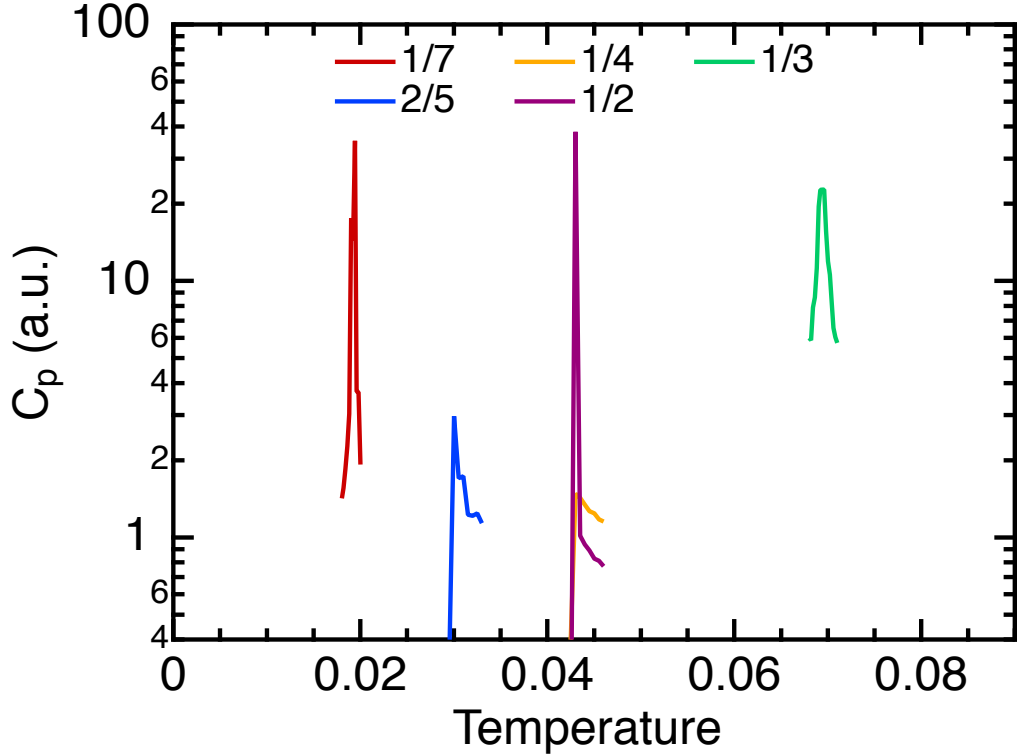

**Extended Data Figure 5 | Transition temperature to the charge-ordered state (simulation).** Transition temperatures are determined from the peak in the temperature dependence of the heat capacity $C_p$ for filling $1/7, 1/4, 1/3, 2/5, 1/2$. Temperature is given in units of the energy $e^2/(4\pi\varepsilon\varepsilon_0 a)$ and $d/a$ is fixed to be 10.

**Extended Data Table 1 | Comparison between model and experiment for the transition temperature of the charge-ordered states at filling 1/7, 1/4, 1/3, 2/5 and 1/2.** The model $T_c$'s are in units of the energy of $e^2/(4\pi\varepsilon\varepsilon_0 a)$ assuming $d/a = 10$. They are evaluated in Kelvins using $\varepsilon = 3.9$, which is chosen to match the highest experimental $T_c$ (at $v = 1/3$). The experimental $T_c$'s are taken from the electron doping side.

|  | $v = 1/7$ | $v = 1/4$ | $v = 1/3$ | $v = 2/5$ | $v = 1/2$ |
|---|---|---|---|---|---|
| $T_c$ (model) | 0.0190 | 0.0430 | 0.0695 | 0.0300 | 0.0430 |
| $T_c$ (K, model) | 10 | 23 | 37 | 16 | 23 |
| $T_c$ (K, exp) | $10 \pm 2$ | $32 \pm 2$ | $37 \pm 1$ | $18 \pm 1$ | $29 \pm 2$ |